\documentclass{article}

\usepackage{lineno,hyperref}
\usepackage{amsmath,amssymb}
\usepackage{graphicx}

\usepackage[utf8]{inputenc}
\usepackage[english]{babel}
\usepackage[T2A]{fontenc}

\usepackage{comment} 

\modulolinenumbers[10]
\modulolinenumbers[1]
\newcommand\nnfootnote[1]{%
  \begin{NoHyper}
  \renewcommand\thefootnote{}\footnote{#1}%
  \addtocounter{footnote}{-1}%
  \end{NoHyper}
}

\begin{document}

\title{Chimera states in ring-star network of Chua circuits}

\author{Sishu Shankar Muni\footnotemark[1] \and Astero Provata \footnotemark[2]} 

{\renewcommand\thefootnote{\fnsymbol{footnote}}%
\footnotetext[1]{School of Fundamental Sciences, Massey University, Palmerston North, New Zealand}
  \footnotetext[2]{Institute of Nanoscience and Nanotechnology,
              National Center for Scientific Research "Demokritos",
              15341 Athens, Greece}}

\maketitle

\begin{abstract}
We investigate the emergence of amplitude and frequency
chimera states in ring-star networks consisting 
of identical Chua circuits connected via nonlocal diffusive, bidirectional coupling. 
We first identify single-well chimera patterns in a ring network under 
nonlocal coupling schemes. 
When a central node is added to the network,
forming a ring-star network, the central node acts as the distributor of information, increasing the chances of synchronization. Numerical simulations show that the radial
coupling strength $k$ between the central and the peripheral nodes acts as 
an order parameter leading from a lower to a higher frequency domain.
The transition between the domains
takes place  for intermediate coupling values, $0.5 <k< 2$,
where the frequency chimera states prevail. The transition region
(width and boundaries)
depends on the Chua oscillator parameters and the network specifics. 
Potential applications of star connectivity can be found 
in the control of Chua networks and in
other coupled chaotic dynamical systems. By adding one central node 
and without further modifications
to the individual network parameters it is possible to entrain 
the system to lower or higher
frequency domains as desired by the particular applications.
 
\nnfootnote{Keywords: Chimera states, Chua circuit, nonlocal diffusive coupling, ring-star network}
\nnfootnote{E-mail addresses: S.S. Muni (\url{s.muni@massey.ac.nz})\\ \hspace*{2.8cm} A. Provata (\url{a.provata@inn.demokritos.gr})}

\end{abstract}

\section{Introduction}
\label{sec:intro}
\par Chimera states are characterized by the coexistence of synchronous and asynchronous areas 
when identical dynamical units are coupled equivalently  in some network topology. 
Most commonly studied are the frequency chimeras, which are distinguished by the 
difference in frequency of the oscillatory elements. Although all oscillators have the
same intrinsic frequency, it is the coupling between them which creates 
a distribution of frequencies in the network, a reason why the chimera states are 
nontrivial and unexpected. Kuramoto and Battogtokh first observed these patterns 
in nonlocally coupled phase oscillators described in \cite{kuramoto:2002,kuramoto:2002a}. 
Following these first reports, chimera states have been studied extensively 
during the past two decades. Researchers found chimera states in many different types 
of dynamical systems, like coupled phase oscillatory flows and discrete maps connected 
in different network topologies and coupling forms,
as described in recent review articles \cite{panaggio:2015,schoell:2016,majhi:2019,oomelchenko:2018}.
Besides the typical chimeras as reported by Kuramoto, where the synchronous and
asynchronous domains remain fixed in space, more complex chimera patterns were recently
reported, such as breathing chimeras, alternating chimeras \cite{Chimera:types}, 
spiral wave chimeras \cite{Chimera:spiral}, 
amplitude chimeras \cite{Chimera:amplitude}, and many more. 

\par Applications of chimera states were first reported in systems of coupled nonlinear
neuronal oscillators \cite{majhi:2019}. 
These oscillators are known to exhibit highly nonlinear behavior and chimera states 
were found in systems that mimic the neuronal activity like the FitzHugh-Nagumo (FHN), 
the Hindmarsh-Rose (HR), and the Hodgkin-Huxley (HH) systems
\cite{omelchenko:2015,Chimera:Hindmarsh,Chimera:twodimneuron,Chimera:LIFneuron}.
 Apart from the numerical studies, experimental evidence of 
chimera states related to neuronal activity has also been reported in \cite{Chimera:experimental},
 in which nine 
FHN oscillators in a ring were considered. Synchronization/coordination aspects 
related to neurological disorders like epilepsy were studied via this system. The results 
indicated that epilepsy is not only a dynamical disease but also a topological disease 
that depends on the type of connection between the neurons. Further indications
on the presence of chimera states at the onset of epileptic seizures are reported
in references~\cite{mormann:2000,mormann:2003,andrzejak:2016}.
It has also been claimed that chimera states are deeply connected with the causes of 
creating various kinds of neuronal diseases like Parkinson's disease, schizophrenia,
 brain tumors, etc \cite{Brain:2006}. 
While performing experiments, Tognoli et al. reported  in reference~\cite{Brain:metastable}
the presence of synchronous and asynchronous activity during left and right finger movements. 
Also in relation to brain activity, chimera states have been associated with the 
unihemispheric sleep pattern 
in aquatic creatures and birds, where they sleep with one eye open leading one half of the 
brain in the synchronous state and the other half in the
 asynchronous \cite{rattenborg:2000,rattenborg:2006,ramlow:2019}. 

\par
Besides experiments related to neuronal activity,  experimental evidence
of chimera states have been reported in diverse disciplines and notably in the domains
of mechanical oscillators \cite{martens:2013}, coupled map lattices \cite{hagerstrom:2012},
nonlocally coupled FitzHugh-Nagumo and Stuart-Landau oscillator circuits
\cite{gambuzza:2014,gambuzza:2020} and chemical oscillators 
\cite{tinsley:2012,wickramasinghe:2013,wilson:2018}. 
Recent numerical evidence indicates that metamaterials
are also a promising domain for chimera applications \cite{hizanidis:2016,hizanidis:2020}.
Apart from the confirmation of the chimera states in experiments and simulations, 
 it still remains an open problem to deal analytically with the mechanism behind 
the formation of these hybrid states and their control,
 making the study of chimeras an active research area.    
  
\par The Chua circuit is considered as the simplest nonlinear circuit to exhibit chaotic 
behavior \cite{Chua}. It is composed of an inductor, two capacitors, a resistor, and a Chua diode. 
The circuit's temporal evolution is described by a three dimensional continuous dynamical system. 
The Chua diode, being a nonlinear local active resistor, is mainly responsible for the chaotic 
behavior of the circuit, which is well known for its double-scroll attractor \cite{Chua:dscroll}. 
Regarding applications, Chua circuits are used in secure communications \cite{Chua:secure}, in improving performance of ultrasonic devices in the
presence of cross talk \cite{Chua:ultrasonic} and in
 the generation of Gaussian, white noise which is essentially useful in many engineering systems \cite{Chua:noise}. Other applications include Avant-Garde music compositions \cite{Chua:music}, storage
of analog patterns and 
managing the problem of handwritten recognition \cite{Chua:recognition}.
Many present-day applications of the Chua circuit are described in reference \cite{Chua:todaybook}.

\par Star networks and ring networks are usually common in social systems, hubs, social networks, computer networks \cite{Star:motiv}. In the latter, the information flow needs to be transmitted securely, else there is a compromise in the security leading to cybercrimes. 
Chua circuits connected in the computer networks,
in particular, can help in achieving secure communications as the
Chua circuits have proven to be useful in many cryptographic systems \cite{Chua:crypto}. 

\par Synchronization of Chua oscillators was studied in star networks 
and their properties were established using different kinds of coupling,
 such as diffusive, conjugate and mean-field couplings \cite{Muni:2018}. The regimes of full synchronization of the Chua oscillator networks were 
mainly studied using these different coupling forms. 
A variety of chimera structures have been reported in the case of  Chua circuits connected in a ring \cite{shepelev:2017}. 
 Chimera states were also observed in the star network consisting of synchronized and desynchronized oscillators in the group. 
In a two-dimensional lattice of Chua oscillators, spiral waves are obtained in \cite{Chua:spiral}. 
The present study is an extension of the previous work \cite{Muni:2018}, in the sense that we are using a
composite connectivity scheme: starting with a ring of Chua circuits with nonlocal connectivity and
common coupling strength $\sigma$ (see \cite{shepelev:2017}),
we apply an additional radial connectivity where every Chua circuit is bidirectionally connected to a 
central node with a variable coupling strength $k$. By varying the values of $k$ and $\sigma$
 we can transit from a pure ring connectivity, when $k=0$ and $\sigma \ne 0$, to  a pure star (central)
connectivity, when $k \ne 0$ and $\sigma =0$. 
We investigate the prevalence of different chimera patterns in this composite connectivity, 
which we refer to as the ring-star Chua network.        

\par In particular, in this study starting from the pure ring connectivity with nonlocal coupling
we recover the chimera structure known as single-well chimera as shown in reference \cite{shepelev:2017}.
To set terminology, here we need to explain the difference between multi-scroll 
attractors, such as the Chua attractors, and multi-level chimeras. A multi-scroll attractor represents the phase space of a  
single chaotic oscillator, whose trajectory circulates in the vicinity of multiple regions,
escaping occasionally from each of these regions to the others \cite{Chua}. 
Chimera states, on the other hand,
are formed in systems of many coupled oscillators, where domains of coherent and incoherent oscillators
are formed. In chimera states the coupled oscillators can be chaotic or simpler limit cycles. When
the coupled elements have complex phase space, multiple domains of synchronized elements are formed,
separated by domains where the oscillators are asynchronous. These are called multi-well chimeras
and consist of regions (wells) of coherent elements with constant common frequencies,
separated by incoherent regions where the oscillators develop different frequencies \cite{shepelev:2017}.
In the present study,  
by turning on the radial coupling $k$, we show that initially, the single-well chimera structures 
persist in the ring-star network and all oscillators have specific low frequencies. 
As $k$ increases, there is an abrupt transition at finite $k$-values, 
where the system passes from the low frequency regime to a high frequency regime, passing by 
the intermediate $k$-region where domains of high and low frequencies co-exist. The $k$-value where the transition occurs depends on the
ring coupling strength $\sigma$. 

\par In the following, when we refer to the exchange of information between nodes
$i$ and $j$ in the system, we mean that at a certain time $t$ the state variables $x_i(t)$ and
$y_i(t)$
of node $i$ receive
and use the values of the state variables $x_j(t-\Delta t)$ and $y_j(t-\Delta t)$ of node $j$ 
in the previous time step
$t-\Delta t$. This must not be confused with the notion of global information, energy or entropy
exchange between the nodes as discussed in the literature of synchronization between interacting
units \cite{pikovski:2001,kouvaris:2010,hou:2011,baptista:2012,soriano:2012,majhi:2016,kundu:2020} and more recently on synchronization in the form of chimera states \cite{bick:2015,deschle:2019}.

\par The transitions from asynchronous patterns to chimera states and to synchronized states are
quantitatively studied with the help of the different synchronization measures \cite{tsigkri:2018}. 
The mean phase velocity \cite{omelchenko:2013} is a common measure considered to demonstrate the 
presence of chimera structures. However, it frequently fails to identify them, 
mainly in the cases of traveling or diffusing chimeras.
 In those cases, there is a need for different measures to be considered, which can work as alternatives
 to the mean phase velocity. Alternative measures  analyzing the relative size of the coherent/incoherent 
domains, the degree of coherence etc were considered in \cite{tsigkri:2018}.
 We establish the chimera kingdom in the ring-star network of Chua circuits using these measures, to 
avoid the problems of pattern displacement in space and for quantitative comparison between the
different chimera morphologies.   

The paper is organized as follows: Section~\ref{sec:model} introduces the Chua circuit and the ring-star network topology. In a separate subsection, \ref{sec:measures}, the various synchronization measures are introduced. Section~\ref{sec:results} discusses the simulation results obtained in the case of 
Chua circuits coupled in a ring geometry for parameter values where single-well chimeras emerge.
Section~\ref{sec:chua-star} is devoted to the 
 ring-star network connectivity. The deformation of the single-well chimeras as a function of the radial coupling $k$ is discussed in this section.
In section \ref{sec:transition} the transition of the system from the lower frequency to the
higher frequency domain is discussed where the frequency chimera states prevail.
  In all cases, alternative synchronization measures are considered for the quantitative study of the
mean phase velocity profiles. In the conclusions the main results of this study are recapitulated and some challenges ahead in analyzing the ring-star network are proposed.

\section{Chua ring-star network model}
\label{sec:model}
\begin{figure}
    \centering
    \includegraphics[width = 0.45\textwidth]{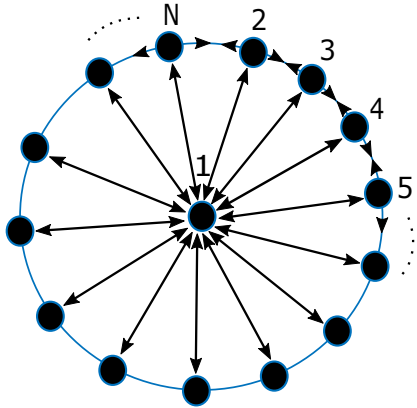}
    \caption{The Chua ring-star network. Here we consider $N=300$ Chua circuits where the central one
 is labeled $i=1$ and the end nodes are labeled from $i=2,\ldots,N$.}
    \label{networkfig1}
\end{figure}

A sketch of the ring-star network is shown in Figure ~\ref{networkfig1}. 
A number of $N$  Chua oscillators are connected in a ring-star network with nonlocal diffusive  coupling. 
Oscillators are indexed as $i=1$ for the central node and $i=2,\ldots,N$ for the peripheral (end) nodes. 
The central node ($i=1$) is connected to all the peripherals with the same coupling strength $k$. Each peripheral oscillator is nonlocally
connected to $R$ nodes to its left and $R$ nodes to its right with common coupling strength $\sigma$ and is also linked to the central node with coupling strength $k$. To enforce uniformity of the end nodes, periodic boundary conditions are considered. 

\par The dynamical equations of the ring-star network are given by Eqs.~\eqref{eq:diffusive} and \eqref{eq:central}. For $i=2,\ldots, N$, the dynamical equations of the end nodes are given by:
\begin{align}
\begin{split}
\label{eq:diffusive}
\dot{x_{i}} &= f_{x} + k(x_{1}-x_{i}) + \frac{\sigma}{2R} \sum_{k=i-R}^{k=i+R}(x_{k}-x_{i}),\\
\dot{y_{i}} &= f_{y}+\frac{\sigma}{2R}\sum_{k=i-R}^{k=i+R}(y_{k}-y_{i}),\\
\dot{z_{i}} &= f_{z}.   
\end{split}
\end{align}
\noindent For $i=1$ (central node) the dynamical equations are: 
\begin{align}
\begin{split}
\label{eq:central}
\dot{x_{1}} &= f_{x} + \sum_{j=1}^{N} k(x_{j} - x_{1}),\\
\dot{y_{1}} &= f_{y},\\
\dot{z_{1}} &= f_{z}. 
   \end{split}
\end{align}
\noindent where
\begin{align*}
f_{x} &= \alpha(y_{i} - x_{i} - (Bx + \frac{1}{2}(A-B)(|x+1| - |x-1|))),\\
f_{y} &= x_{i} - y_{i} + z_{i},\\
f_{z} &= -\beta y_{i}.   
\end{align*}
\noindent with periodic boundary conditions:
\begin{align*} 
x_{i+N}(t) &= x_{i}(t),\\
y_{i+N}(t) &= y_{i}(t),\\
z_{i+N}(t) &= z_{i}(t).
\end{align*}
\noindent for $i=2,3,\ldots,N$. Following references \cite{Muni:2018,shepelev:2017}, we have used coupling only in the $x$ and $y$-variables and not in the $z$-variable of the Chua coupled elements.
Similar coupling only via one variable is used in reference~\cite{sharma:2011} for coupled R{\"o}{ss}ler oscillators.
\par 
From the interaction scheme, it is now clear that oscillators $i=2,\ldots,N$ exchange information
via their $x-$ and $y-$variable with $2R$ neighbors symmetrically set around $i$, while the central unit $i=1$
exchanges information with all other units $j=2,\ldots,N$ (also via their $x-$ and $y-$variables only). Due to the Euler integration scheme used, 
the variables $x_i(t),y_i(t),z_i(t)$ are updated using the values  
$x_i(t-\Delta t),y_i(t-\Delta t),z_i(t-\Delta t)$ at the previous time step, as also
stated in the Introduction.
\par As working parameter set the following values are used throughout this study: 
The parameters of the identical Chua circuits are set to $A=-1.143$, $B =-0.714$,
$\alpha = 9.4$ and $\beta = 14.28$ in order to keep the circuit in the oscillatory,
double-scroll regime. 
The system size is set to $N=300$ and the coupling
range to $R=100$. The rest of the parameters, the coupling strength $\sigma$ 
between the peripheral nodes, and the coupling strength $k$ between the central node 
and the peripheral ones are varied to explore their influence in the 
network synchronization patterns. 

\subsection{Synchronization measures}
\label{sec:measures}
As discussed in the Introduction, the mean phase velocity or average frequency $\omega$ is 
a valuable measure to quantify the synchronization of the oscillators \cite{omelchenko:2013}. 
For the $i$-th oscillator, the mean phase velocity is denoted by $\omega_{i}$. 
For a large computational time interval $T$, $\omega_i$ expresses the number of  times the 
variable $x_{i}$ crosses a certain fixed constant value, say $c$. If the variable $x_{i}$ 
crosses the constant $c$, $M_{i}$ times with positive slope, 
then the mean phase velocity of the $i$-th oscillator 
is calculated as :
\begin{equation}
\omega_{i} = \frac{2\pi M_{i}}{T}=2\pi f_i .
\label{omegaidefn}
\end{equation}
\noindent The positive slope considered in the counting of $M_i$ in Eq.~\eqref{omegaidefn}
is needed to avoid double counting the number of periods calculated
within the time interval $T$. The quantity $f_i$ denotes the average frequency, 
which differs from the 
mean phase velocity by a factor $2\pi$. Due to this simple relation, 
in the following the terms ``mean
frequency'' and ``mean phase velocity'' will be used interchangeably.

\par Chimera states are characterized by the difference in frequency of the 
identical oscillatory circuit elements. 
The coupling is responsible for the change in frequency in some oscillators. 
This is the reason why the chimera states are so nontrivial and unexpected. 
Different synchronization measures come to play as additional quantitative indices
when inconclusive information is conveyed by the mean phase velocity. 
Such synchronization measures were discussed in \cite{tsigkri:2018}. We complement our work with the computation of the relative size of the incoherent and coherent parts denoted by $r_{\textit{incoh}}$ and $r_{\textit{coh}}$, respectively. Let us denote by $\omega_{\text{coh}}$ the common  mean phase velocity
of the coherent elements, by $\omega_{i}$ the mean phase velocity of the $i$-th oscillator and by 
$N$ the number of oscillators considered. 
The quantity $r_{\textit{coh}}$ is defined as :
\begin{equation}
r_{\textit{coh}} = \frac{1}{N} \sum_{i=1}^{N} \chi(A_1)
\label{rcohdefn}
\end{equation}
where $\chi$ is a step function which returns $1$ if $A_1$ is positive and returns $0$ if $A_1$ is negative. 
$A_1$ is defined as:
\begin{equation}
A_1 =   \|   \omega_{i} - \omega_{\text{coh}} \| - \epsilon
     \label{defnA1}
\end{equation}
In definition ~\eqref{defnA1} a small tolerance $\epsilon$ is added 
in order to take into account the fluctuations at the coherent level while computing 
$r_{\textit{coh}}$ according to Eq.~\eqref{rcohdefn}. 

\par Similarly, the quantity $r_{\textit{incoh}}$
 is defined as :
\begin{equation}
r_{\textit{incoh}} = \frac{1}{N} \sum_{i=1}^{N} \chi(A_2)
\label{rincohdefn}
\end{equation}
where 
\begin{equation}
A_2 = \begin{cases}
     \omega_{i} - \omega_{\text{coh}} - \epsilon , & \omega_{\text{coh}} < \omega_{\text{incoh}}\,, \\
     \omega_{\text{coh}} - \omega_{i} - \epsilon , & \omega_{\text{coh}} > \omega_{\text{incoh}} 
     \end{cases}
     \label{defnA2}
\end{equation}

\noindent  Note that in the definition ~\eqref{defnA2} two cases are considered. That is because 
one needs to cover both cases: when the $\omega_{\rm coh}$ coincides with the minimum frequency in
the system (upper case in Eq.~\eqref{defnA2}) and  when the $\omega_{\rm coh}$ coincides with the maximum frequency in
the system (lower case in Eq.~\eqref{defnA2}). Both cases have been recorded in the literature on 
chimera states, see references~\cite{omelchenko:2015,tsigkri:2018,ulonska:2016}.
In our simulations, the tolerance level $\epsilon $ is fixed to be $0.2\Delta \omega$, where
$\Delta \omega =\omega_{\rm max}-\omega_{\rm min}$ is the difference between the maximum and
minimum $\omega_i$ values in the system. 

\section{Chimera states in  Chua ring networks}
\label{sec:results}
\begin{figure*}[hb!]
    \centering
    \includegraphics[width = 1\textwidth]{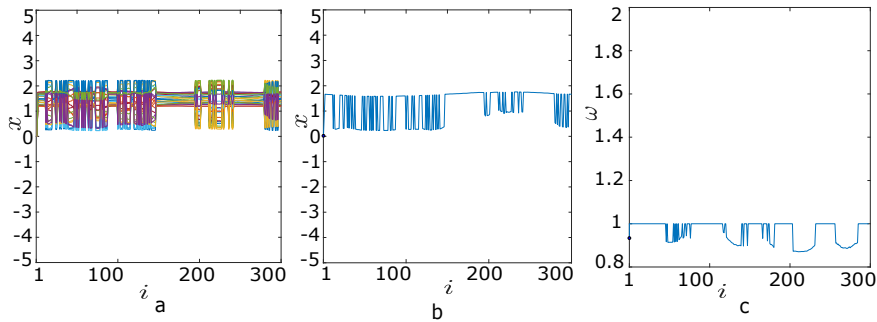}
    \caption{Ring network: Single-well chimera structures and measures for $\sigma = 0.75, k = 0$ 
with nonlocal diffusive coupling.
a) 25 snapshots of the $x_i$-variables at time intervals of 40 units, b) typical single snapshot of the
$x_i$-variables and c) mean phase velocities.
Chua circuit parameters are $A = -1.143,\>\> B = -0.714,\>\> \alpha = 9.4, \>\>
 \beta = 14.28$ and network parameters
are $N=300$ and $R=100$. 
}
    \label{fig:01}
\end{figure*}

\par In this section, we study the case of a  ring network where 
each element is a Chua circuit nonlocally linked to its
neighbors, as was proposed and studied by Shepelev et al. in reference \cite{shepelev:2017}. 
Following \cite{shepelev:2017} we consider here
the formation of chimeras in the case of ring connectivity with nonlocal diffusing coupling.
To avoid unnecessary complexity which arises in the parameter regions where the double-well chimeras
prevail, we restrict ourselves in the parameter regions where
only  single-well chimera states are observed. 

\par  
Starting with a ring network of $N=300$
Chua oscillators with nonlocal diffusive coupling, we record single chimera states 
for different values of the network parameters (work by Shepelev et al. \cite{shepelev:2017}).
As an exemplary case, in Fig.~\ref{fig:01} we record the spatial and temporal behavior
of the network for coupling parameters $\sigma =0.75$ and $k=0$. Panel~\ref{fig:01}a depicts the spatial
profile of the network at 25 snapshots at time intervals of 40 units.
 This figure depicts a single-well
chimera state, where domains of alternating oscillatory properties are formed. 
Having started with initial conditions,
$(x(t=0),y(t=0),z(t=0))$, 
randomly distributed
 in the interval 

\noindent
$\left[ 0< x(t=0)<1,\>\> 0< y(t=0)<1,\>\> 0< z(t=0) <1\right] $,

\noindent the system
remains always in the positive side of the axes (single-well) and all elements
oscillate around the value $x=1.5$,
but amplitude variations are observed in the different domains formed.
For clarity, a single
snapshot is presented in panel~\ref{fig:01}b. 
The mean phase velocity, panel~\ref{fig:01}c, clearly identifies the domains where oscillators act
coherently, and the incoherent domains.

\par In the next section, \ref{sec:chua-star}, we introduce a central node to the system, creating the
ring-star network, and we investigate the system's response by varying  the radial coupling
strength $k$. We study the system response in the case of single-well  chimeras, using as coupling constant
 parameter $\sigma =0.75$ and variable $k$, as discussed above. 

\section{Chimera states in ring-star Chua networks}
\label{sec:chua-star}

\begin{figure*}
    \centering
    \includegraphics[width = 1\textwidth]{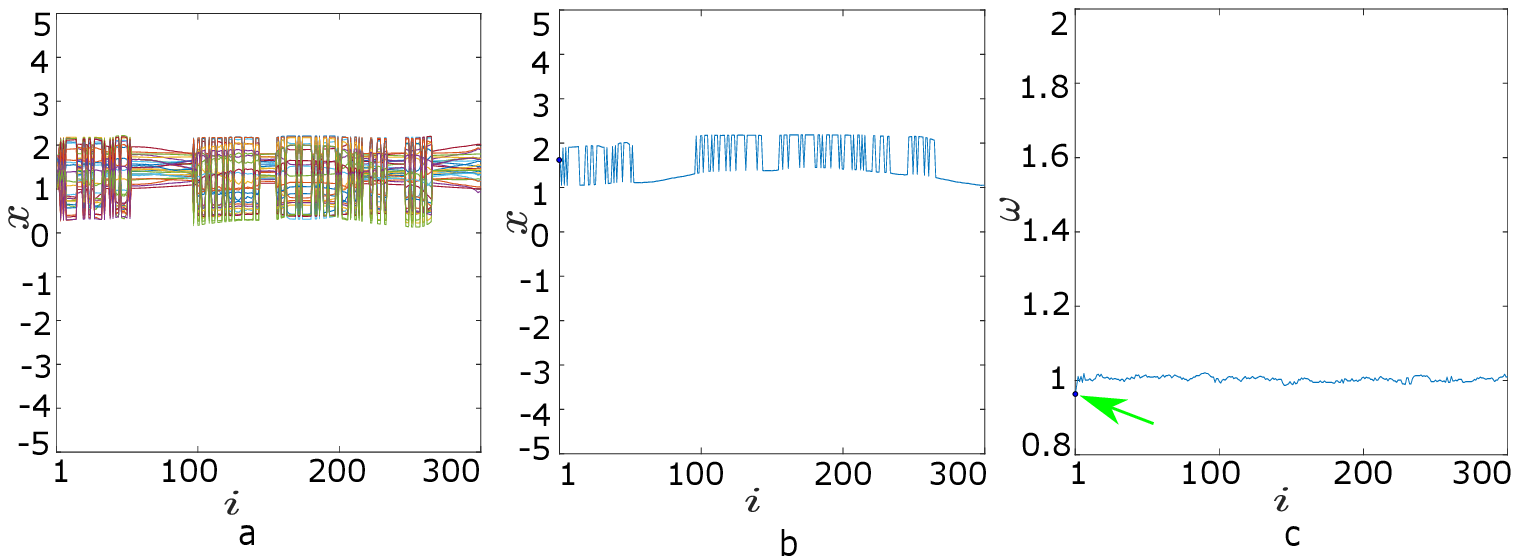}
    \caption{Ring-star network: Single-well chimera structures and measures for
 $\sigma = 0.75, k = 0.25$ with nonlocal diffusive coupling.
a) 25 snapshots of the $x_i$-variables at time intervals of 40 units, b) typical single snapshot of the
$x_i$-variables and c) mean phase velocities. The arrow in panel c) indicates the mean phase velocity of the central node.
All other parameters as in Fig.\ref{fig:01}.
}
    \label{fig:a1}
\end{figure*} 
\begin{figure*}
    \centering
    \includegraphics[width = 1\textwidth]{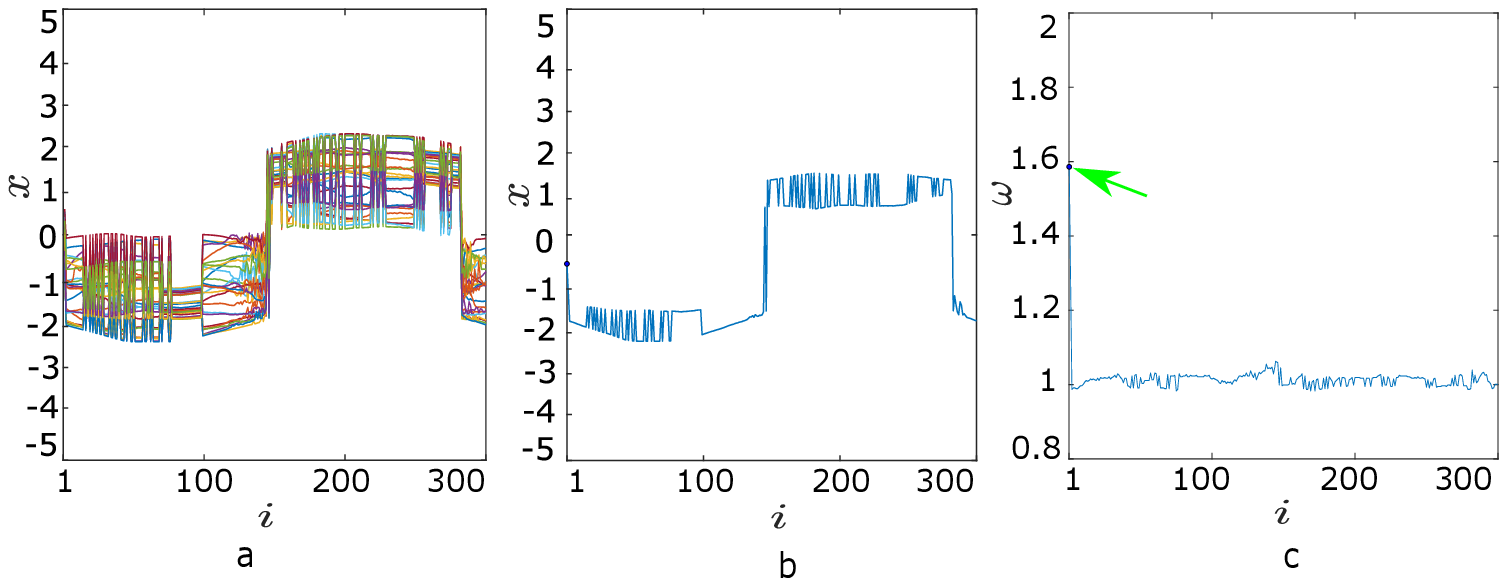}
    \caption{Ring-star network: Double-well chimera structures and measures for 
$\sigma = 0.75, k = 1$ with nonlocal diffusive coupling.
a) 25 snapshots of the $x_i-$variable at time intervals of 40 units, b) typical single snapshot of the
$x_i-$variable and c) mean phase velocities. The arrow in panel c) indicates the mean phase velocity of the central node.
All other parameters as in Fig.\ref{fig:01}.
}
    \label{fig:002}
\end{figure*}
\begin{figure*}
    \centering
    \includegraphics[width = 1\textwidth]{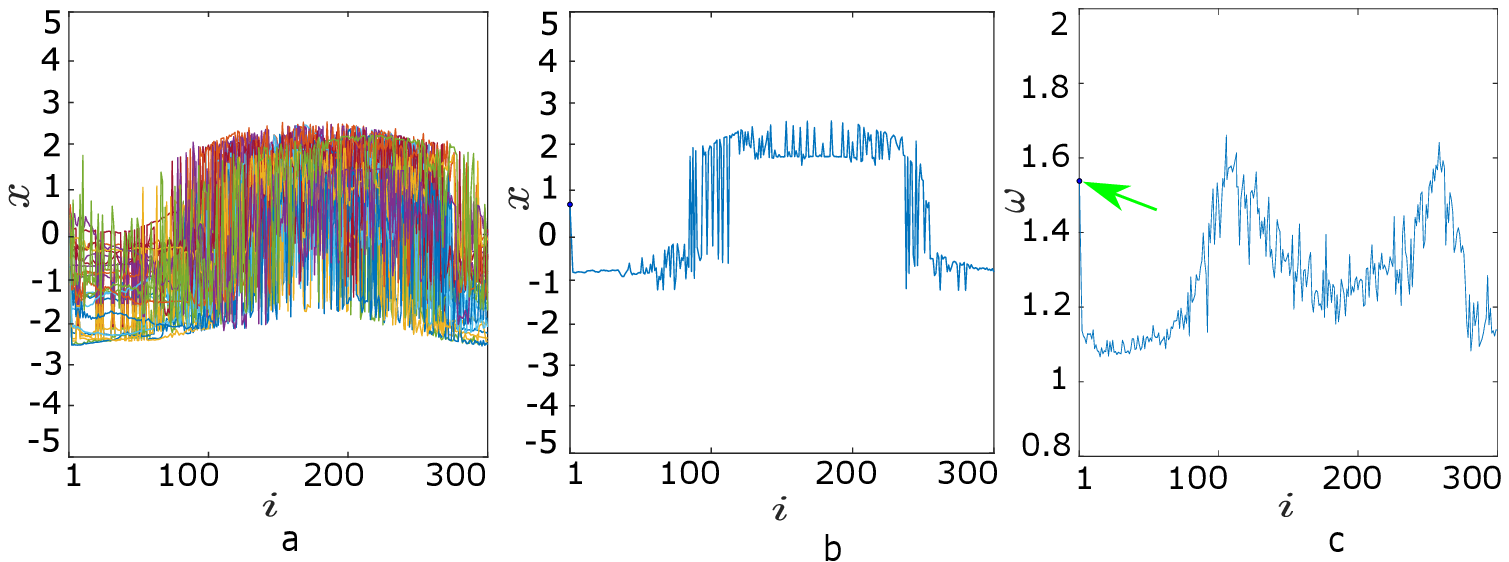}
    \caption{Ring-star network: Double-well chimera structures and measures for $\sigma = 0.75, k = 1.3$ with nonlocal diffusive coupling.
a) 25 snapshots of the $x_i-$variables at time intervals of 40 units, b) typical single snapshot of the
$x_i-$variables and c) mean phase velocities. The arrow in panel c) indicates the mean phase velocity 
of the central node.
All other parameters as in Fig.\ref{fig:01}.
}
    \label{fig:003}
\end{figure*}
\begin{figure*}
    \centering
    \includegraphics[width = 1\textwidth]{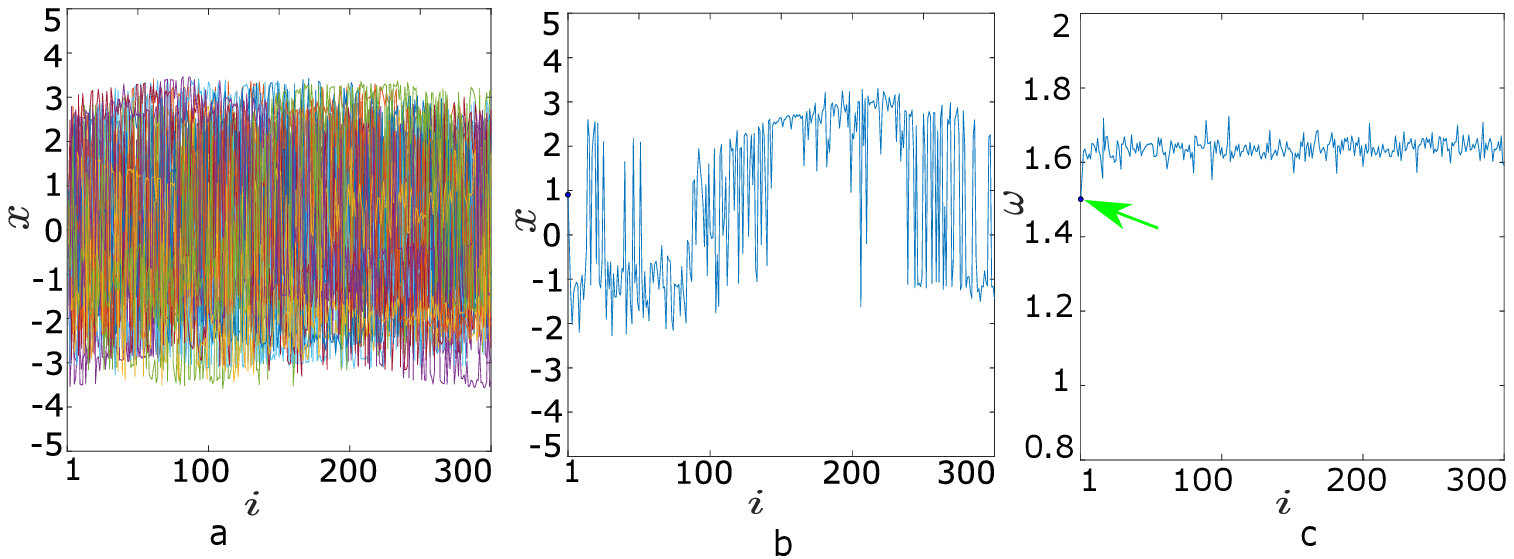}
    \caption{Ring-star network: Double-well chimera structures and measures for $\sigma = 0.75, k = 2.5$ with nonlocal diffusive coupling.
a) 25 snapshots of the $x_i-$variables at time intervals of 40 units, b) typical single snapshot of the
$x_i-$variables and c) mean phase velocities. The arrow in panel c) indicates the mean phase velocity 
of the central node.
All other parameters as in Fig.\ref{fig:01}.
}
    \label{fig:lastk}
\end{figure*}

We now extend the Chua ring network model by adding 
a central node in the center of the
ring which is linked equally to all the external nodes. We consider the mixed dynamics of the system
and record its transition between chimera states and the full synchronization regimes.
 Simulations were carried out by considering random initial conditions for $x,y,z$ state variables in the interval $[0,1]$, as in previous section.

\par To investigate the influence of the central node in the Chua ring network we gradually
 vary the coupling
strength $k$ between the central node and the peripheral ones. All 
 parameters of the identical Chua circuits are kept to the working parameter set, while the network
contains $N=300$ nodes and each Chua oscillator is connected to $R=100$ neighbors to the left and
$R=100$ neighbors to the right.
The ring coupling strength is fixed to $\sigma = 0.75$. 
[The pure-ring network briefly discussed in the previous section
is equivalent to the case of $k=0$ (no central node), while
the ring-star network is realized when $k \neq 0$.] 

\par The central node of the network plays a double role: First it receives ``information'' from the
peripheral nodes and integrates it forming its own dynamics. Second, it redistributes the obtained
information to the peripheral nodes, in such a way that each peripheral node receives information
about the average (mean-field) dynamics of the ensemble of all peripheral nodes. Therefore, the central
node acts as a modulator of the local dynamics using information over the ensemble dynamics.
Based on this view of the system,
we ask the question: How does the strength of the central coupling $k$ influences
the distribution of information and the overall synchronization properties of the network? To
answer this question we performed numerical simulations using the same parameters as in the
case of the single-well chimera
and varied the central coupling parameter in the range $0 \le k \le 4$.
\par 
We provide below some examples of the modifications which take place when the central coupling
strength $k$ becomes nonzero. When a small deviation is applied leading from 
 the ring network, $k=0$, to the ring-star network, $k=0.25$, as in Fig.~\ref{fig:a1}, we observe that the single-well chimera persists:  the $x$ state variable keeps
oscillating in the positive part of the axis and does not transverse below $x=0$.
For this low $k-$value, all oscillators have very similar frequencies, as can be seen in Fig.~\ref{fig:a1}c. The green arrow in panel (c) points to the mean phase velocity $\omega_1$ of the 
central node, which is slightly lower than the rest of the elements.
Increasing the radial strength to $k=1$ in Fig.~\ref{fig:002} and $k=1.3$ in Fig.~\ref{fig:003},
the single-well chimeras change to double-well ones. In both cases two domains of oscillators 
are formed: one domain 
where the $x$-variables oscillate in the positive axis and one in the negative axis.
Related to the mean phase velocity values, in the case of $k=1$, Fig.~\ref{fig:002}c, all peripheral
nodes have the same mean phase velocity, while the central node has increased its 
mean phase velocity,
 $\omega_1 \sim 1.6$,
indicating a tendency of the system to transit to higher frequencies 
(see position 
of the green arrow in Fig.~\ref{fig:002}c). 
Increasing further the $k$-values, see the case $k=1.3$ in Fig.~\ref{fig:003}c,
the nodes occupying the transition regions, between the negative and the positive $x-$value domains,
 follow the central node (see position 
of the green arrow) and attain in their turn higher mean phase velocities. 
With a further increase in the radial coupling strength $k$, for example $k=2.5$, 
we observe that the $x$ state variables still traverse both the positive and negative part 
of the axis as in Fig.~\ref{fig:lastk}. Furthermore, in panel (c) we observe that all mean-phase velocities 
have increased as compared to lower values of $k$.

\par In the next section, sec.~\ref{sec:transition}, we study how the transition from the the single to
double-well chimera takes place as we gradually increase the radial coupling strength $k$.

\section{Dynamics with increase in the radial coupling strength}
\label{sec:transition}
We analyze here the dynamics of the ring-star Chua network as the 
 coupling strength $k$ (the ray coupling strength) increases between $0 \le k \le 4$.
The ring coupling strength is fixed as $\sigma = 0.75$. We study the behavior of the mean phase velocity
of the 
central node, $\omega_{\rm central}$, of the coherent oscillators, $\omega_{\rm coh}$, 
and of the ``leader''
incoherent oscillator, $\omega_{\rm leader}$. From Fig.~\ref{fig:003} 
in the previous section, we record in the incoherent regions  a 
continuous distribution of
frequencies and not a single one.
In these regions we call ``leader'' the oscillator which demonstrates the maximum mean phase velocity, which,
for this reason is called $\omega_{\rm leader}$. Note that there can be more than one leaders in the system,
one for each incoherent region, as Fig.~\ref{fig:003} indicates. In a way, the leader oscillators can be
considered as the ones which lead the deviations from coherence,  while the difference
$\Delta\omega =  \omega_{\rm leader}-\omega_{\rm coh} $ is indicative
of the total incoherence in the system.

\begin{figure}[h!]
\begin{center}
\hspace*{2.5cm}
\includegraphics{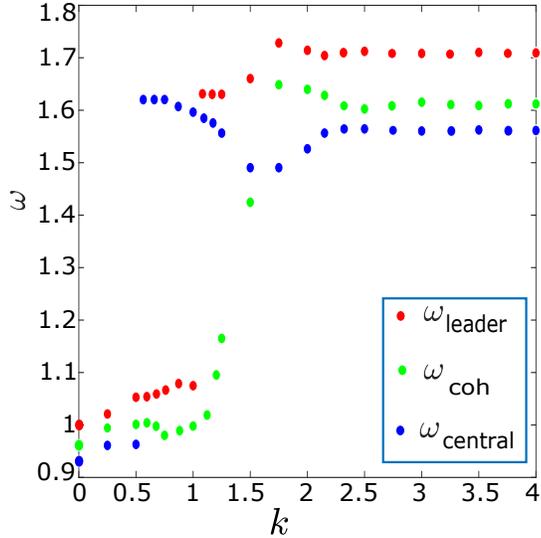}
\vspace*{-6cm}\caption{Mean phase velocities as a function of the coupling strength $k$.
The mean phase velocities of the coherent nodes, $\omega_{\rm coh}$, are marked in green color,
 of the central node, $\omega_{\rm central}$, is marked in blue color,
while the maximum mean phase velocity in the incoherent regions, $\omega_{\rm leader}$, is marked in red color.
All other parameters as in Fig.\ref{fig:01}.}
\label{fig:004}
\end{center}
\end{figure}

\par In Fig.~\ref{fig:004} we plot the 
 mean phase velocities of the central node $\omega_{\rm central}$ (blue color),
of the coherent nodes $\omega_{\rm coh}$ (green color), and
of the leader incoherent nodes $\omega_{\rm leader}$ (red color). Initial conditions were chosen randomly in all simulations within the positive interval, $[0,1]$, for the $x,y,z$ state variables.
 This figure indicates the presence of
a phase transition taking place in the parameter region $0.5 \le k_{\rm trans} \le 1$.
In particular, for small values of the radial coupling, $k \le 0.5$, all oscillators present
similar $\omega$-values, around $\omega \sim 1$. In this region ($k \le 0.5$),
 the central oscillator, $i=1$,  has the smallest frequency, slightly 
below the coherent ones, while the leaders have frequencies slightly above the coherent.  
As $k$ increases above 0.5 the abrupt transition occurs. First the central nodes 
double (almost) their mean phase 
velocity which becomes close to 1.6, while the
rest of the oscillators remain close to the values $\omega \sim 1$. This behavior holds
in the intermediate coupling region, $0.5 \le k \le 1.0$ 
(for the parameter values $\sigma =0.75$, $N=300 $ oscillators
and $R=100$ neighbors). Above this transition region, and for $k > 1$, the coherent and 
incoherent nodes also increase gradually their mean phase velocities, which also attain values around
$\omega \sim 1.6$. In particular, for values $1.25 < k <1.5$, 
the central node frequency is located
between the coherent and the leader ones.
When $k$ reaches strengths $> 2.0$ the oscillator regions stabilize and the system attains
constant frequencies, $\omega_{\rm central}\sim 1.55$, $\omega_{\rm coh}\sim 1.6$
and $\omega_{\rm leader}\sim 1.72$ independent of $k$. 
The above discussion tells us that the frequency
``chimera kingdom'' characterized by considerable differences
in the frequencies between coherent and incoherent domains
is established for $k$-parameter values in the transition region $1.0 < k < 2.0$. 
\par
\begin{figure}[h!]
\begin{center}
\hspace*{-0.8cm}\includegraphics[width = 0.85\textwidth]{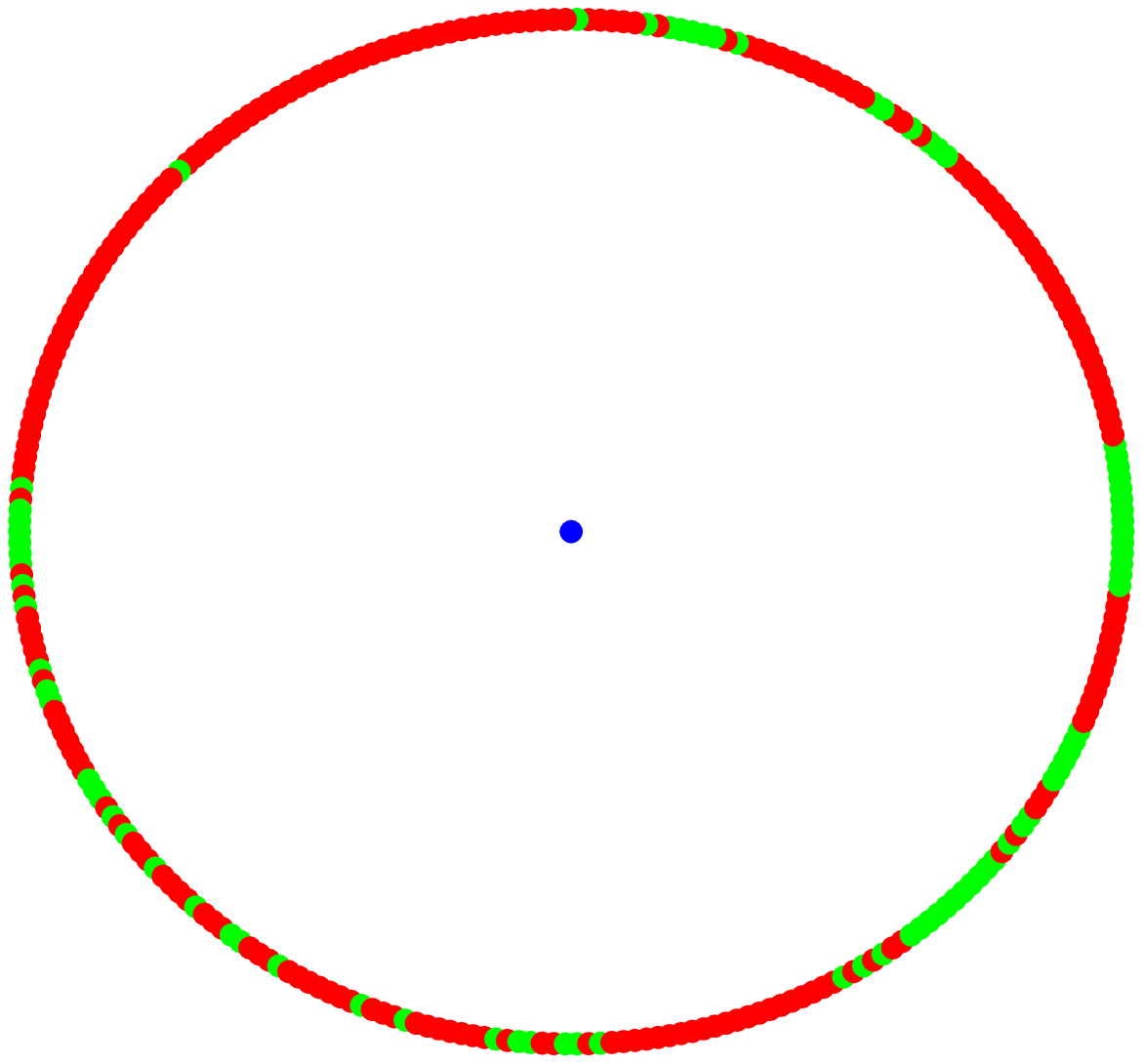}
\caption{Coherence circle plot. The coherent nodes are marked with green, incoherent nodes are marked with red color. The number of nodes depicted here is $N =300$. The central node is marked with blue color. The coupling strengths are fixed as: $k=1$ and $ \sigma = 0.75$. All other parameters as in Fig.\ref{fig:01}.}
\label{fig:coherence_circle}
\end{center}
\end{figure}

 Figure~\ref{fig:coherence_circle} shows the ring-star network in action. 
We represent the central (blue) node, coherent (green) and incoherent (red) nodes for $k=1, \> \sigma = 0.75$ in the ring-star network with different colors. As we see from the figure, the coherent nodes
are not isolated but form clusters in the ring. At the same instant the incoherent nodes are most abundant. 
\par
\begin{figure}[h!]
\begin{center}
\includegraphics{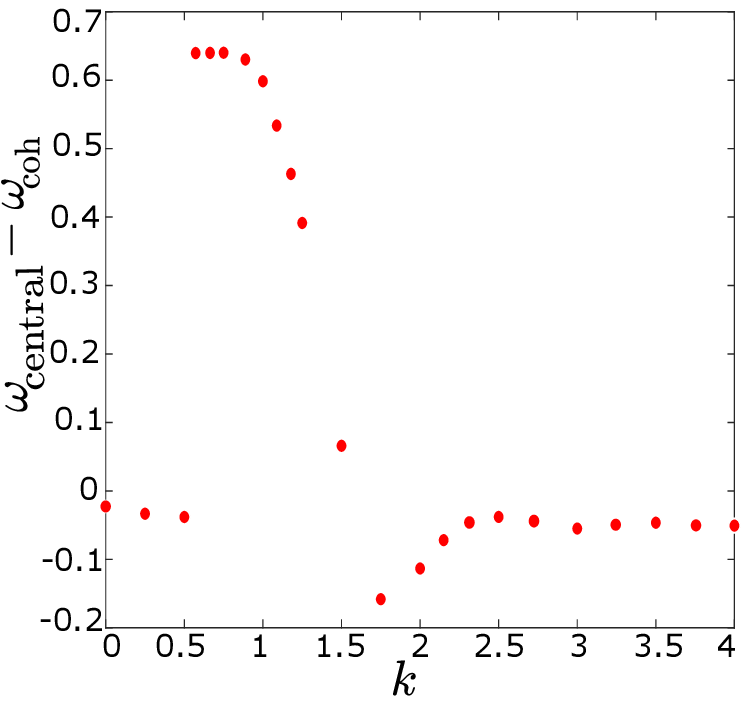}
\caption{The difference $\omega_{\rm central}-\omega_{\rm coh}$ as a function of the
radial coupling strength $k$. Ring coupling strength $\sigma = 0.75$ and other parameters as in Fig.~\ref{fig:002}.
}
\label{fig:005}
\end{center}
\end{figure}

\begin{figure}[h!]
\begin{center}
\hspace*{2.2cm}
\includegraphics{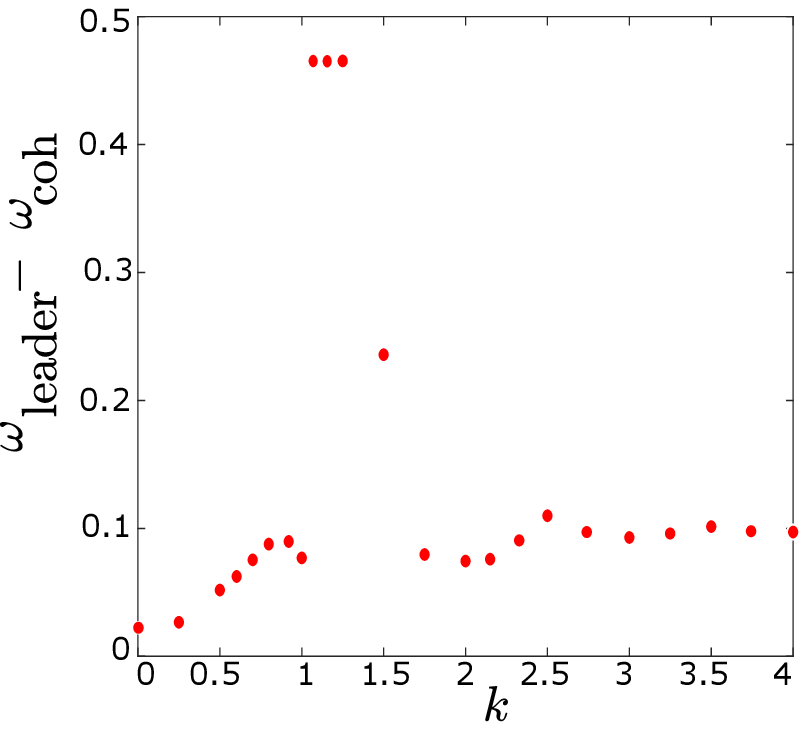}
\caption{The difference $\omega_{\rm leader}-\omega_{\rm coh}$ as a function of the
radial coupling strength $k$. Ring coupling strength $\sigma = 0.75$ and other parameters as in Fig.~\ref{fig:002}.
}
\label{fig:006}
\end{center}
\end{figure}

 
The transition shown in Fig.~\ref{fig:004} is corroborated by the plots depicting
 $\omega_{\rm central}-\omega_{\rm coh}$ and $\omega_{\rm leader}-\omega_{\rm coh}$ in
Figs.~\ref{fig:005} and ~\ref{fig:006}, respectively. 

\par Namely, in Fig.~\ref{fig:005} we note that the values $\omega_{\rm central}-\omega_{\rm coh}$
remain slightly below 0, for $k<0.5$ indicating that the central node has constantly lower frequency
than the coherent nodes. Above $k>2$, the same behavior persists. In the intermediate region,
$0.5 < k < 2.0$, first the central node acquires abruptly very high frequencies for $0.5 <k< 1$, and
later on its frequency gradually decreases to stabilize slightly below the frequency of the coherent nodes.
We recall that, in all cases,
 the central node has the same characteristics (parameters) as all other nodes
in the system. 
Concerning the leader nodes, in Fig.~\ref{fig:006} the values $\omega_{\rm leader}-\omega_{\rm coh}$
remain above 0 for $k<1.0$, demonstrating a gradual increase as a function of $k$ in this region.
They also demonstrate a delay in the
transition with respect to the central node. The leaders are entrained to transit for $k\sim 1.25$
while the coherent ones reach the higher frequency domain after $k \sim 1.5$.
Above $k>2$, the leaders keep a constant small mean phase velocity difference from the coherent nodes
of the order $\sim 0.05$.

\par
\begin{figure}[htbp!]
\begin{center}
\hspace*{1.3cm}
\includegraphics[width=0.65\textwidth]{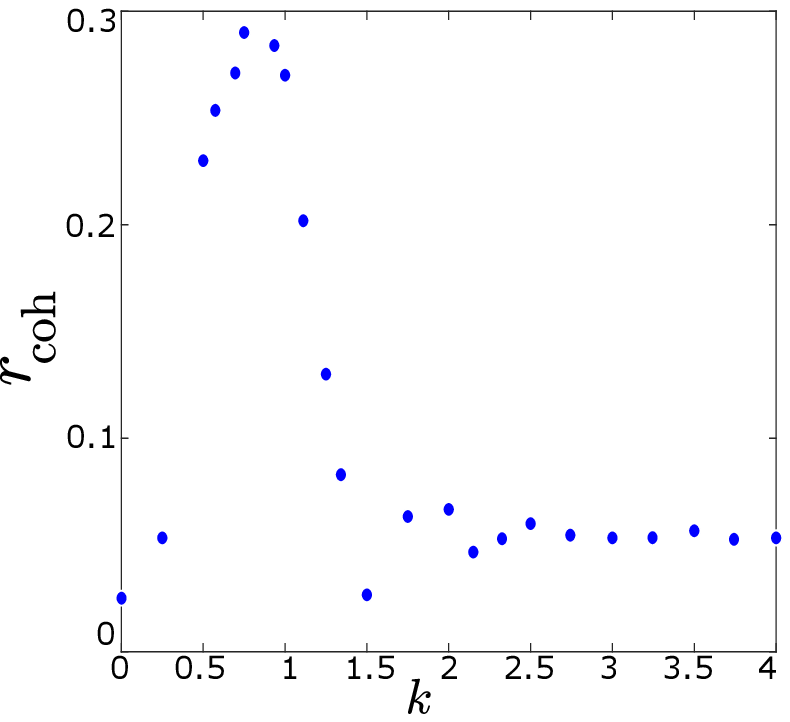} 
\caption{Ratio of coherent elements as a function of the central coupling range  $k$.
All other parameters are as in Fig.~\ref{fig:01}.
}
\label{fig:0010}
\end{center}
\end{figure}

\begin{figure}[htbp!]
\begin{center}
\hspace*{1.3cm}\includegraphics[width=0.65\textwidth]{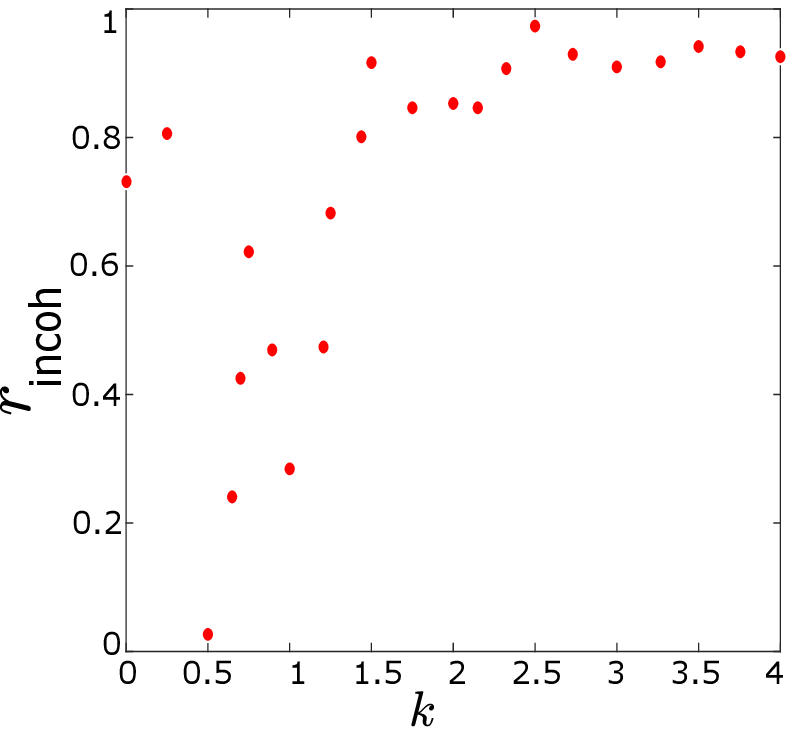}
\caption{Ratio of incoherent elements as a function of the central coupling range  $k$.
All other parameters are as in Fig.~\ref{fig:01}.
}
\label{fig:0011}
\end{center}
\end{figure}

 To study further the inhomogeneity of frequencies in the system we compute the ratio 
of coherent, $r_{\rm coh}$
and incoherent nodes $r_{\rm incoh}$ as a function of $k$, excluding the central one.
The ratios of coherent  and  incoherent elements were calculated using Eqs.~\eqref{rcohdefn} and \eqref{rincohdefn} and the
results are depicted in Figs.~\ref{fig:0010} and \ref{fig:0011}, respectively. 
We have a similar picture of the transition from small  to large frequency values
with increasing the strength $k$ of radial coupling. For small ($k<0.5$) and large ($k>2$) 
radial coupling strengths the ratio of coherent elements stays low, $r_{\rm coh}\sim 0.1$ and the
ratio of incoherent stays large, $r_{\rm incoh}\sim 0.9=1-r_{\rm coh}$. In the intermediate region,
$0.5 < k <2$ a transitive behavior is recorded, where the coherent ratio increases, while the
incoherent one decreases. This reorganization of the system taking place in the
intermediate $k$ regions where the frequency chimera states prevail, marks the passage
from the lower to the  higher frequency domain.

\par As a general conclusion, in the transition between low and high frequencies, first the central node
makes the jump to the higher frequencies at $k\sim 0.5$ entraining the rest of the nodes.
Following the central node the incoherent nodes are entrained. Their leaders 
make the transition at $k \sim 1$, while the coherent nodes attain the transition at $k \sim 1.5$.
Note that the present values are indicative and hold for the working parameter set. For different 
parameters $(\sigma , N, R)$, the transition values as well as the transition regions are expected to vary depending
on these parameters.

\section{Conclusions}
\par We have studied a ring network of Chua circuits, equipped with a central node which serves to
redistribute to the peripheral nodes information about the mean field state of all nodes. 
For small values of the radial coupling strengths single-well amplitude chimeras are observed. 
At intermediate radial couplings a transition region is observed where the 
frequency chimeras prevail with a large difference in the frequencies between coherent and incoherent
nodes. This region mediates the transition between the lower and higher frequency domains.
For large radial coupling strengths, the system attains the higher frequency domain and keeps
constant mean phase velocities and ratios of coherent to incoherent nodes, independent of
 the radial coupling range.
The frequency chimera kingdom is established for the intermediate radial couplings $k$-values, 
as evidenced by the plots of all different synchronization measures.

\par The above results have potential applications in the control of Chua networks as well as
other coupled chaotic dynamical systems. By just
adding one central node, identical to all peripheral ones, and without further 
modifications
to the individual oscillators or to the network parameters, 
it is possible to entrain the system to lower or higher
frequency domains as desired by the particular applications by only adjusting the radial coupling.
We must stress here that the transition described above is an example of transitions
taking place in nonequilibrium systems (nonequilibrium transitions); 
the Chua system \eqref{eq:diffusive} is a characteristic example of such systems since it presents chaotic, nonconservative dynamics
\cite{Chua,Chua:dscroll,Chua:secure,Chua:todaybook,Chua:spiral}.
This transition cannot be directly
related to the known phase transitions in equilibrium systems at criticality, 
such as the Ising model phase transitions (see reference \cite{stanley:1971}).

\par For future studies, it would be interesting to
 understand how the dynamics of the Chua network changes with different coupling forms such as conjugate coupling or mean-field coupling, or by strengthening the role of the central node
and endowing it with interactions to the
peripheral nodes using all three $x$, $y$ and $z$ variables.  Transitions in
different network types may also be considered as, for example, in the case of a
 2D lattice of Chua oscillators equipped with a central element, or 
extensions to Chua circuits in multilayer arrangements.
\par A different study concerns the connection between star and ring-star networks. In order to achieve a complete star network (central node connected to peripheral nodes and no connection in between the peripheral nodes), mathematically we need to fix a finite value of $k$ while letting $\sigma \rightarrow 0$. It would be interesting to account for chimera states and potential transitions in this limiting $\sigma$ case and to investigate the link to the phenomenon of Remote Synchronization (RS)
\cite{sharma:2011,bergner:2012},
a nontrivial phenomenon in star networks, where the peripheral oscillators synchronize (without being directly linked), while the central, relay node remains asynchronous. 
\section*{Acknowledgements}
This work was supported by computational time granted from the Greek Research \& Technology Network
(GRNET) in the National HPC facility - ARIS - under project CoBrain4 (project ID: PR007011).



\end{document}